\documentclass[10pt]{iopart}

\usepackage{graphicx,
            amsmath,
            gensymb,
            hyperref,
            color,
            comment}
\newcommand{\change}[1]{#1}
\usepackage[margin=1in,footskip=0.25in]{geometry}

\begin{document}

\title[]{Associative memories using complex-valued Hopfield networks based on spin-torque oscillator arrays}

\author{Nitin~Prasad$^{1,2}$, Prashansa~Mukim$^{1,2}$, Advait~Madhavan$^{1,3}$, and Mark~D.~Stiles$^{4}$}
\address{$^{1}$ Associate, Physical Measurement Laboratory, National Institute of Standards and Technology, Gaithersburg, MD, USA}
\address{$^{2}$ Department of Chemistry and Biochemistry, University of Maryland, College Park, MD, USA}
\address{$^{3}$ The Institute for Research in Electronics and Applied Physics, University of Maryland, College Park, MD, USA}
\address{$^{4}$ Physical Measurement Laboratory, National Institute of Standards and Technology, Gaithersburg, MD, USA}
\ead{nitin.prasad@nist.gov}
\vspace{10pt}


\begin{abstract}
Simulations of complex-valued Hopfield networks based on spin-torque oscillators can recover phase-encoded images. Sequences of memristor-augmented inverters provide tunable delay elements that implement complex weights by phase shifting the oscillatory output of the oscillators. Pseudo-inverse training suffices to store at least 12 images in a set of 192 oscillators, representing $16\times 12$ pixel images. The energy required to recover an image depends on the desired error level. For the oscillators and circuitry considered here, 5~\% root mean square deviations from the ideal image require approximately 5~{\textmu}s and consume roughly 130~nJ. Simulations show that the network functions well when the resonant frequency of the oscillators can be tuned to have a fractional spread less than $10^{-3}$, depending on the strength of the feedback.
\end{abstract}

%
%
%
%
%

\section{Introduction}

The need to realize energy-efficient, high-density, and high-speed computing is driven by the explosive emergence of big data~\cite{villars2011big}. One problem with scaling traditional processors to meet this need is that they require shuttling data between memory and the computational processors, a problem referred to as the von Neumann bottleneck. Inspired by biological neural networks, in-memory computing has advanced as an approach to minimize this bottleneck in the last decade~\cite{stanley2017computing, indiveri2015memory}. Novel nanoscale devices offer unique pathways to complement existing complimentary metal-oxide-semiconductor (CMOS) technologies in creating artificial neural networks~\cite{kim2015nvm, wang2017memristors, torrejon2017neuromorphic} in which memory is embedded in the computational engines. 

Neural activity in the brain has inspired a variety of ways to encode information to make processing it more efficient. Mimicking spiking behavior has led to spike-based neural networks that encode and process information based on the strength, rate, and timing of spiking signals~\cite{gautrais1998rate, brette2015philosophy}. Similarly, spontaneous neuronal oscillations observed in collections of biological neurons have inspired oscillatory neural computers that encode and process information as relative frequencies and phases ~\cite{hoppensteadt1999oscillatory, levine1999oscillations}. The parallels between associative learning observed in the brain~\cite{thompson1997associative} and spontaneous synchronization observed in weakly coupled oscillators~\cite{izhikevich2006weakly} has led to proposals for artificial associative memories based on synchronized oscillators~\cite{hoppensteadt1999oscillatory, nikonov2015coupled}. 

One approach to synchronized-oscillator-based artificial associative memories is to physically realize binary Hopfield networks~\cite{hoppensteadt1999oscillatory, abbott1990network}. 
Traditional Hopfield networks are {\em auto-associative} because they can retrieve learned data when presented with a defective or partial piece of data. These networks are recurrent, all-to-all connected, artificial neural networks in which the neurons are binary threshold units~\cite{hopfield1982neural}. The neuron outputs are determined by applying thresholds to the sum of the incoming weighted feedback connections from all other neurons. Information is encoded in the real-valued weights forming these feedback connections. When presented with an incomplete or a noisy version of one of the stored vectors encoded in these weights, the Hopfield network retrieves the corresponding stored vector by performing energy minimization on the functional defined by the network training. Such networks address the von Neumann bottleneck when the weights are stored in proximity to the neurons.
In oscillator-based binary Hopfield networks, the information to be processed is encoded as the relative phases of the oscillators (in-phase or out-of-phase) and the memory used in the processing is stored in the synaptic weights.

There have been several advances in the computational power of Hopfield networks since their original proposal~\cite{hopfield1982neural}. These include advances in training algorithms~\cite{storkey1999basins} and energy functionals~\cite{krotov2016dense, demircigil2017model} that allow for greater information storage~\cite{ramsauer2021hopfield}. Binary Hopfield networks have also been extended to store and retrieve continuous- or quasi-continuous-valued information, such as grayscale images~\cite{rieger1990storing, zurada1996generalized, jankowski1996complex}. This is achieved by using multilevel real activations~\cite{rieger1990storing, zurada1996generalized} or by using complex-valued signum activations~\cite{jankowski1996complex}. As we demonstrate in this work, the latter approach can be mapped to a network of weakly coupled oscillators.

Physically realizing oscillators can be done with CMOS technology alone~\cite{best2007phase}, but such oscillators are not particularly compact. Alternative oscillators for associative memories can be fabricated from emerging nanoscale devices such as spintronic oscillators~\cite{nikonov2015coupled}, vanadium-dioxide oscillators~\cite{parihar2017vertex, nunez2021oscillatory} and opto-electronic oscillators~\cite{farhat85optical, jang88associative}. In this work, we focus on auto-associative memories built using particular spintronic oscillators, spin-torque oscillators realized with magnetic tunnel junctions, although many of the arguments presented here are general and can easily be extended to other oscillator-based systems.

Spin-torque oscillators based on magnetic tunnel junctions (MTJs) consist of a stack of two ferromagnetic layers separated by a thin tunnel barrier insulator~\cite{yuasa2004giant}. Two effects make MTJs useful as the foundation for making oscillators. The first is that due to the tunneling magnetoresistance effect, the device resistance depends on the relative orientation of the magnetizations in the two magnetic layers so that as the magnetization precesses, the resistance and hence either current through or voltage across the MTJ oscillates. The resistance varies roughly as $R(\theta)=R_\mathrm{P}+(R_\mathrm{AP}-R_\mathrm{P})(1-\cos\theta)/2$, where $\theta$ is the angle between the magnetization directions, $R_\mathrm{P}$ is the low resistance found for the parallel alignment, and $R_\mathrm{AP}$ is the larger value for antiparallel alignment.
The second effect that makes MTJs useful as oscillators is that the magnetization can be dynamically excited by passing a current through the tunnel junction. The carriers tunneling through the insulator are spin-polarized by one layer and perturb the magnetic orientation in the other layer through a process called spin-transfer torque~\cite{ralph2008spin}. In most applications, the magnetization of one of the ferromagnetic layers is {\em pinned} and that of the other, {\em free} layer responds to the  current passing through the MTJ. In magnetic random-access memory (MRAM)~\cite{apalkov2013spin}, spin transfer torques due to current pulses are used to switch the magnetization. In spintronic oscillators~\cite{chen2016spintorque}, a DC current excites gigahertz dynamical oscillations that can be detected by the resulting time-varying resistance, which for MTJs is due to the tunneling magnetoresistance.


For memory applications, tunnel junctions are formed so that the parallel and antiparallel configurations of the free layer are energetically favored. Any perturbations of these states give rise to oscillations in the magnetization of the free layer, which decay after the external perturbations are removed~\cite{liu2011ferromagnetic}. A spin-torque oscillator can be realized by tailoring the geometry of an MTJ in a way that sustained microwave-frequency oscillations can be induced in the MTJ when the damping force is balanced by the bias current-induced spin-transfer torque~\cite{kim2012spin}. When external small current perturbations are applied alongside the bias currents, the spin-torque oscillators can lock to these perturbations, provided that the driving frequencies are sufficiently close to the natural frequencies of the spin-torque oscillators~\cite{rippard2005injection,tamaru2015extremely}. In this paper, we consider the well-studied vortex-based spin-torque oscillator~\cite{pribiag2007magnetic}. Such oscillators have been used in other neuromorphic computing implementations such as reservoir computing~\cite{torrejon2017neuromorphic}, synchronizing an array to external sources for speech formant recognition~\cite{romera2018vowel}, and synchronizing an array to external sources to implement synaptic weights~\cite{leroux2021radio}. An array of related oscillators based on metallic nanocontacts have been synchronized for norm computation~\cite{koo2020distance}.

Computing with oscillators typically involves coupling them together with controllable interactions. There is significant research into physically coupling spintronic oscillators through magnetostatic or exchange interactions~\cite{kaka2005mutual,chen2016spintorque,zahedinejad2020two}, which could be compact and energy efficient. However, experiments are just beginning to demonstrate control~\cite{zahedinejad2021memristive} over the oscillators and their coupling. Applications of coupled spin-torque oscillators so far have used electrical coupling~\cite{romera2018vowel,koo2020distance}, taking advantage of tunneling magnetoresistance and spin-transfer torques. The goal of the present work is to present an application based on electrically coupled oscillators and energy-efficient electrical circuits that enable this approach. These circuits allow control of both the amplitude and the phase of the interaction.

In addition to the spin-torque oscillator, efficient implementations of Hopfield networks require local memory. We implement such memory with memristors, which are programmable two-terminal nonvolatile resistors with history-dependent resistances~\cite{wang2020resistive}. Memristors have been made from several classes of materials including transition metal oxides like VO$_\mathrm{x}$~\cite{yi2018biological}, TiO$_\mathrm{x}$~\cite{choi2005resistive}, and HfO$_\mathrm{x}$~\cite{govoreanu201110}; perovskites~\cite{xiao2016energy} and chalcogenides~\cite{li2013ultrafast}. Different mechanisms are responsible for resistive switching behavior in different materials and are described in~\cite{wang2020resistive}. Their appeal is in their potential ability to address the von Neumann bottleneck by providing dense non-volatile local memory. When used as digital cross-point memories, memristors have been integrated in the back-end-of-the-line (BEOL) in various fabrication facilities~\cite{golonzka2019non,jain201913}. In an analog context, memristors are commonly used in a crossbar array where the resistances encode the elements of a matrix to enable matrix-vector multiplication~\cite{burr2015large,prezioso2015training}, the workhorse operation for machine learning acceleration. Here, local memory is used to store the control currents for oscillators and the delay and scale parameters in the implementation of the synapses.

We model a continuous-time Hopfield network using specific models for these two emerging technologies. We implement the neurons with spin-torque oscillators based on CoFeB/MgO/CoFeB magnetic tunnel junctions~\cite{pribiag2007magnetic} and use generic resistive memristor models~\cite{yu2021compute} to program the delays and scaling in the feedback network. In Sec.~\ref{sec:STNOcHopfield}, we describe a schematic structure that implements such a network with spin-torque oscillators functioning as neurons and memristors providing the local control of the delay and scale circuitry that operate as weights in the feedback network. We simulate a network with $N=192$ neurons with $192\times 191$ complex weights (self-feedback is omitted), which are used to store twelve 16$\times$12-pixel images. The simulations reported in Sec.~\ref{sec:simulations} demonstrate retrieval of these images from distorted versions in 5~{\textmu}s while consuming 130~nJ of energy. Sec.~\ref{sec:circuits} details the CMOS circuitry that implements this network for a particular spin-torque oscillator model. Memristors, in which the resistance can be varied, store the delay times and the scaling of the complex weights. For the system size considered here, the implementation of the complex weights consumes comparable energy to the spin-torque oscillators. For larger systems, we expect the energy implementing the complex weights to dominate because the number of weights scales as $N^2$ relative to the oscillators and the amplifiers, which scale as $N$. The details of the model used for the oscillators in the network simulations is described in \ref{sec:appA}. In Sec.~\ref{sec:discussion}, we discuss some of the limits of our zero-temperature model for the spin-torque oscillators, the idealized memristor models, and the progress that is needed in device fabrication to fully realize this implementation of a Hopfield network.

\section{Spin-torque-oscillator-based complex Hopfield networks}
\label{sec:STNOcHopfield}

The Hopfield network proposed in Ref.~\cite{jankowski1996complex} achieves multivalued information storage by replacing the binary state of each neuron and real-valued weights used in the original Hopfield network~\cite{hopfield1982neural} with continuous unit-circle complex states for each neuron and complex-valued weights. In this complex Hopfield network~\cite{jankowski1996complex}, time evolution occurs asynchronously and in discrete time steps, so that the state of each node is maintained on the complex unit circle by choosing the activation functions to be complex-valued signum functions, $\text{sgn}(z)=z/|z|$. 

\begin{figure}
    \centering
    \includegraphics[width=\textwidth]{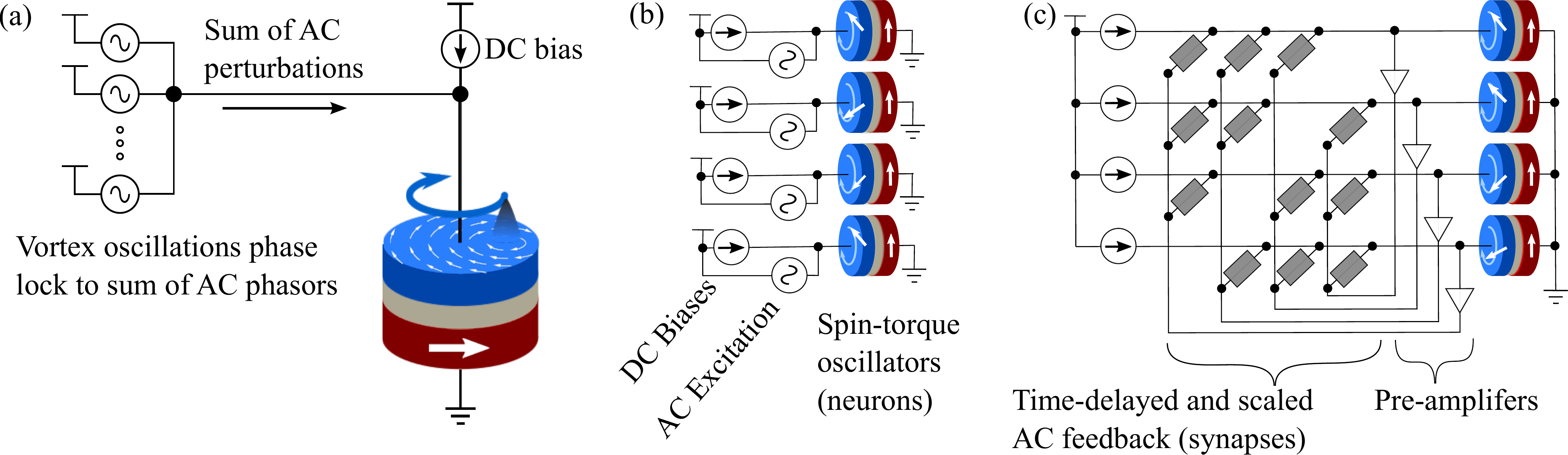}
    \caption{(a) A vortex-based spin-torque oscillator, which has the required characteristics of a complex Hopfield neuron.  (b) The {\em preparation} stage of the retrieval cycle of a complex Hopfield network. The relative phases of the oscillator array are set using external AC signals to reflect the value encoded in the query image. (c) The {\em recognition} stage. The external AC signals are disconnected, and the complex Hopfield feedback network is engaged. The gray rectangles represent the delay-and-scale networks, and the triangles the amplifiers}
    \label{fig:concept_circuits}
\end{figure}

Here, we implement a complex Hopfield network with a physical model of frequency-synchronized coupled oscillators, specifically vortex-based spin-torque oscillators, as depicted in Fig.~\ref{fig:concept_circuits}(a). These oscillators are well characterized~\cite{pribiag2007magnetic} and have been used in several experimental implementations of neuromorphic computing~\cite{torrejon2017neuromorphic,romera2018vowel}. 
As in the approach of Ref.~\cite{jankowski1996complex}, the frequencies of the oscillators are tuned to be close enough that the information is encoded in the relative phases of the oscillators. The absolute phase does not play a role. However, unlike Ref.~\cite{jankowski1996complex}, both the phase and time evolution are continuous because we are modeling the physical behavior of the oscillators and the electronic circuits. That is, the evolution of the network is described by differential equations rather than discrete steps in time. During the time evolution of the oscillators, their phases advance or retard in a continuous manner due to coupling to the other oscillators. Their instantaneous outputs are passed through the complex weights which scale and introduce additional delays to the oscillator outputs. The outputs of the complex weights are then fed back through small AC currents, which together with the nonlinearity of the oscillators determine the time evolution of their phases. The phases evolve because the feedback causes small changes in amplitude that shift the frequency due to the nonlinearity, in turn shifting the phase. We neglect the small changes in the amplitudes except for their effect on the frequencies and phases. Eventually, each oscillator synchronizes to the AC input from the other oscillators with a modified phase.

Training a coupled-oscillator-based complex Hopfield network to store a set of vectors requires setting the complex weights in the feedback loop. Then, the trained network is used to recover stored images from distorted versions of those images. We start by describing the retrieval of stored images using an already trained complex Hopfield oscillator network and return to a discussion of training at the end of this section. The retrieval procedure consists of two stages, a {\em preparation} stage, which uses the circuit schematically illustrated in Fig.~\ref{fig:concept_circuits}(b), and a {\em recognition} stage, which uses the circuit schematically illustrated in Fig.~\ref{fig:concept_circuits}(c). In the preparation stage, the oscillators are uncoupled and driven by external AC signals that individually set the phase of each oscillator. The relative phases of these input signals encode the distorted query vector to be presented to the network. In the recognition phase, the outputs of the oscillators are fed back to the other oscillators to adjust their relative phases to one of the stored images.

\begin{figure}
    \centering
    \includegraphics[width=\textwidth]{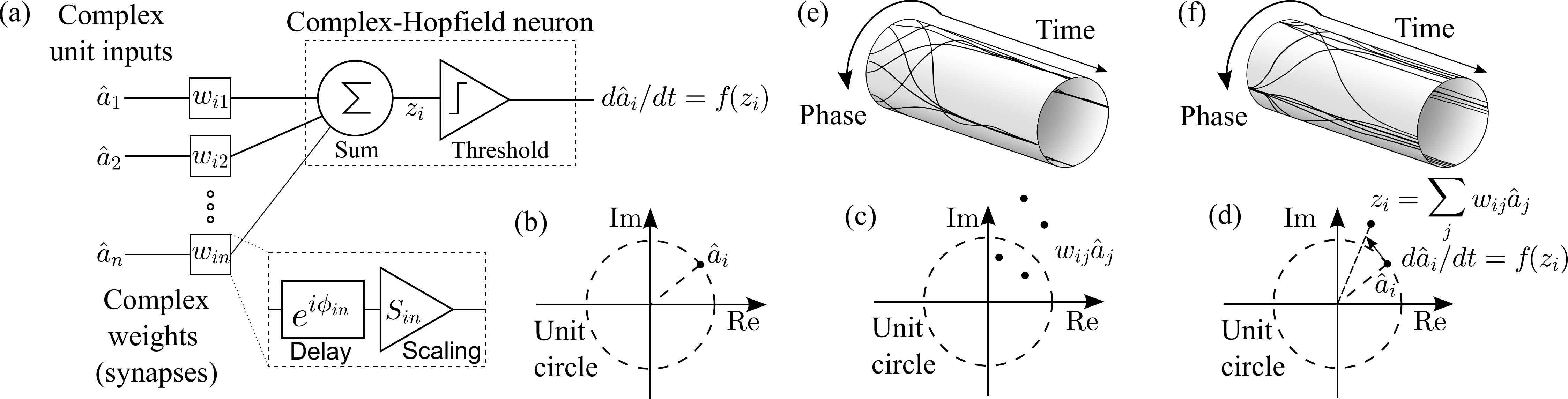}
    \caption{Functionality of a complex Hopfield neuron. (a) Schematic describing the flow of phases and amplitudes.  The phase of oscillator $i$ is represented by the unit normalized complex number $\hat{a}_i$ as in (b). The feedback for each oscillator's phase is the sum of the phases of all other oscillators, each multiplied by a complex weight. Individual terms are shown in (c) and the sum is given by $\hat{z}_i$ in (d). The time evolution of the oscillator phase is shown as a function $f(z_i)$ of the input to the oscillator. (e) Schematic time evolution of the relative phases of the oscillators in the preparation stage and (f) recognition stage. The phases of oscillators vary around the circumference and the time varies along the height of the cylinder. For clarity, this illustrative example shows the phases for images encoded by only four values. In (e) the initial phases are random, the oscillators are uncoupled, and the phases settle into a state representing a query image. Since some of the pixels in the query image are incorrect, in (f) they evolve to the correct phases for the closest stored image under the influence of the feedback.}
    \label{fig:concept}
\end{figure}

Ignoring small changes in the amplitude of the oscillations, the voltage oscillations of the $i$th oscillator can be parameterized by a unit complex number $\hat{a}_i$, as shown in Fig.~\ref{fig:concept}(b). The phase of the AC voltage oscillation corresponds to the angle of $\hat{a}_i$ with $|\hat{a}_i|=1$ set by the oscillation amplitude. A schematic of the time evolution of the relative phases of the spin-torque oscillators as they phase lock to the input signals during the preparation stage is depicted in Fig.~\ref{fig:concept}(e). The phases are plotted on a cylindrical manifold with the angle on the circular axis representing the phase of an oscillator and the dimension along the length of the cylinder representing time. Each solid black line on the cylinders in Fig.~\ref{fig:concept}(e) represents the time evolution of the phase of one of the oscillators as it starts from a random phase and stabilizes into one of the four states in the query image. For illustrative purposes in this example, the nominal information is encoded by relative phases of 0, $\pi/2$, $\pi$ and $3\pi/2$. The stored images in this example have these discrete phases in each of the pixels. The noisy query image also has these values but some of them are incorrect with respect to the stored image. 

After each oscillator individually phase locks to its input signal with a phase encoding the trial image, the external AC drives are turned off while the feedback circuit is turned on to start the recognition stage. The feedback circuit used in the recognition phase, implemented by delay-and-scaling networks, is schematically illustrated in  Fig.~\ref{fig:concept_circuits}(c) and conceptually in Fig.~\ref{fig:concept}(a). The feedback to each oscillator from the other oscillators is an AC current determined by the \change{phases of the} AC voltage outputs of the other oscillators and the appropriate pairwise weights. When a mixture of AC perturbations is applied to an oscillator close to its natural frequency, the oscillator output evolves toward phase-locking to the sum of these AC perturbations. The output voltage of the $j$th oscillator in the network, represented by $\hat{a}_j$, is multiplied by a set of complex-valued synaptic weights $w_{ij}$, which connect the output of the $j$th oscillator to the input of all other oscillators. The resulting products for all $j\neq i$, represented by dots in Fig.~\ref{fig:concept}(c), are fed into the $i$th complex Hopfield neuron which first adds up all the incoming weighted inputs to compute $z_i = \sum_j w_{ij}\hat{a}_j$. \change{Information is encoded only in the phases of the oscillators and not in their amplitudes. However, as is the case in any Hopfield network, both phases and amplitudes of the weights in the feedback network are important as they capture the interference between the many stored vectors}. 

The feedback causes the phases of the oscillators to adjust relative to each other and to settle into the relative phases corresponding to the stored image that is closest to the noisy input image.The oscillator state $\hat{a}_i$ evolves continuously towards the instantaneous values of $\text{sgn}(z_i)$ with a rate $d\hat{a_i}/dt$ dependent on  $\hat{a}_i$ and $z_i$, as depicted in Fig.~\ref{fig:concept}(d). The evolution of $\hat{a}_i$ does not have a simple functional form as discussed in \ref{sec:appA} but is determined by the equation of motion of the oscillator. Finally, in steady state, $\hat{a}_i = \text{sgn}(z_i)$ for all $i$. A schematic illustration of this evolution is given in Fig.~\ref{fig:concept}(f). Here, the phases each start from the chosen discrete value in the query image and evolve toward the closest stored image. After allowing sufficient time for the network to stabilize, the relative phases settle to a steady state that encodes the recalled vector. Pixels that start with the wrong value change significantly as they approach the correct values and pixels that start with the correct values are pulled slightly away from those values before they settle back into the correct values.

The feedback circuit used in the recognition phase is shown schematically in Fig.~\ref{fig:concept_circuits}(c). The feedback path consists of a preamplifier, used to boost the weak microwave output voltages of the spin-torque oscillators, and complex-valued weight elements. The DC biases across each oscillator have been tuned using memristors so that each of the oscillators has the same frequency. Because all the oscillators in the network run at the same frequency, the complex weight elements can be implemented using a time-delay network followed by an amplifier network as in Fig.~\ref{fig:concept}(a). Both the delays and scalings are programmed by tuning the resistances of memristors, see Sec.~\ref{sec:circuits}.  The outputs of these weight elements inject small-signal AC currents (small relative to the DC current) into the oscillators that get added to the constant DC bias currents. 

\begin{figure}
    \centering
    \includegraphics[width=1.0\textwidth]{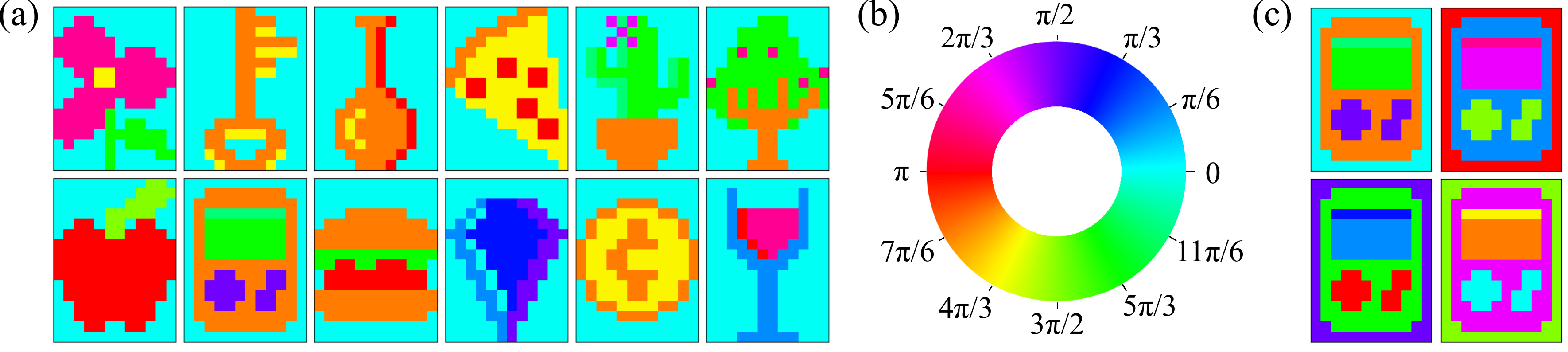}
    \caption{(a) Image dataset used in this study. Each image has 16$\times$12 pixels. The colors used are fully saturated and can be represented using the color wheel in (b). (b) Color wheel with phase labels corresponding to the 12-color palette used in the dataset in (a). (c) Equivalent representations of an image that vary by absolute phases.}
    \label{fig:dataset}
\end{figure}

We demonstrate the two-stage retrieval process by simulating an array of $N = 192$ vortex-based spin-torque oscillators. The complex weight elements encode $K = 12$ images, each with $16\times12$ pixels, shown in Fig.~\ref{fig:dataset}(a). These images are fully saturated in color and the color of each pixel corresponds to one of the twelve discretized phases that are equivalent to twelve particular states of the color wheel in Fig.~\ref{fig:dataset}(b). The chosen discrete phases are indicated on the color wheel. The circular nature of the color wheel allows us to naturally map the colors to the periodic phases of the oscillator. Although such a dataset with periodic pixel values is naturally suited to be coded on a complex Hopfield network, demonstrations of linear grey-scale image learning and retrieval have also been demonstrated on software-based complex Hopfield networks~\cite{jankowski1996complex, muezzinoglu2003new}. In such linear scale mappings, however, an additional error may be introduced during the retrieval process if the extreme values of the linear scale are mapped to adjacent discretized phases. Note that in the phase encoding used here, only the relative phases of the oscillators encode information. An absolute phase added to all oscillators does not affect the dynamics. For example, all four images in Fig.~\ref{fig:dataset}(c), which differ from each other by an absolute phase, represent the same image.

The complex-valued weights $w_{ij}$ are set so that when the network of oscillators is connected in the feedback configuration, the original vectors stored on the oscillator array correspond to fixed points of the oscillator dynamics. A wide selection of learning rules is available. Non-iterative rules such as Hebbian~\cite{hopfield1982neural} and pseudo-inverse~\cite{kanter1987associative} learning rules provide single-shot offline training methods when provided with the entire dataset. On the other hand, iterative training rules such as contrastive divergence~\cite{movellan1991contrastive} and Storkey~\cite{storkey1999basins} learning rules provide iterative methods of learning starting from an initial guess solution. Here, we first use a pseudo-inverse learning rule for offline training, which can store correlated vectors more efficiently than the Hebbian rule~\cite{kanter1987associative} and explore iterative training rules at the end of Sec.~\ref{sec:simulations}. The pseudo-inverse weights $w_{ij}$ are given by
\begin{equation}
    w_{ij} = \frac{1}{N}\sum_{k,l = 1}^{K}\hat{x}_i^{k} [C^{-1}]^{kl}\left(\hat{x}_j^{l}\right)^* \quad\text{and} \quad C^{kl} = \frac{1}{N} \sum_{i=1}^{N} \left({\hat{x}_i}^{k}\right)^*{\hat{x}_i}^{l}.
\end{equation}
Here, ${\hat x}_i^k$ are unit complex-numbers encoding the $i$th pixel of the $k$th image to be stored on the oscillator array. Ref.~\cite{kanter1987associative} showed that altering the values of the diagonal elements of the weight matrix $w_{ii}$ does not affect the fixed points of the oscillator dynamics. Reducing the magnitude~\cite{gorodnichy1999optimal} of the diagonal weights, $w_{ii}$, or setting them to zero~\cite{kanter1987associative} have been proposed as ways to improve convergence towards fixed points. We choose to set $w_{ii} = 0$ and avoid self-feedback.

When the oscillator network is connected in the feedback configuration using the weights, $w_{ij}$, the phases, described by ${\hat a_i}$, continuously evolve toward the steady-state phase-locked condition, as shown in Fig.~\ref{fig:concept}(d). As they evolve, the network stabilizes into a local minimum of the energy $E = -\sum_{i,j=1}^{N} \hat{a}_i^* w_{ij} \hat{a}_j$.

The complex-valued weights can be implemented using a delay-and-scale network. The details of a complex weight implementation using CMOS circuits augmented with memristors to store the weights locally are described in Sec.~\ref{sec:circuits}. In Sec.~\ref{sec:simulations}, we assume that these are ideal delay and scale elements. We capture the non-linear dynamics of a vortex-based spin torque oscillator using the model described in \ref{sec:appA}. Using this oscillator model along with the ideal delay elements, we perform system-level simulations, described in Sec.~\ref{sec:simulations}, to study the two stages of the inference cycle in an offline-trained network that stores the images in Fig.~\ref{fig:dataset}(a).


\begin{figure}
    \centering
    \includegraphics[width=0.8\textwidth]{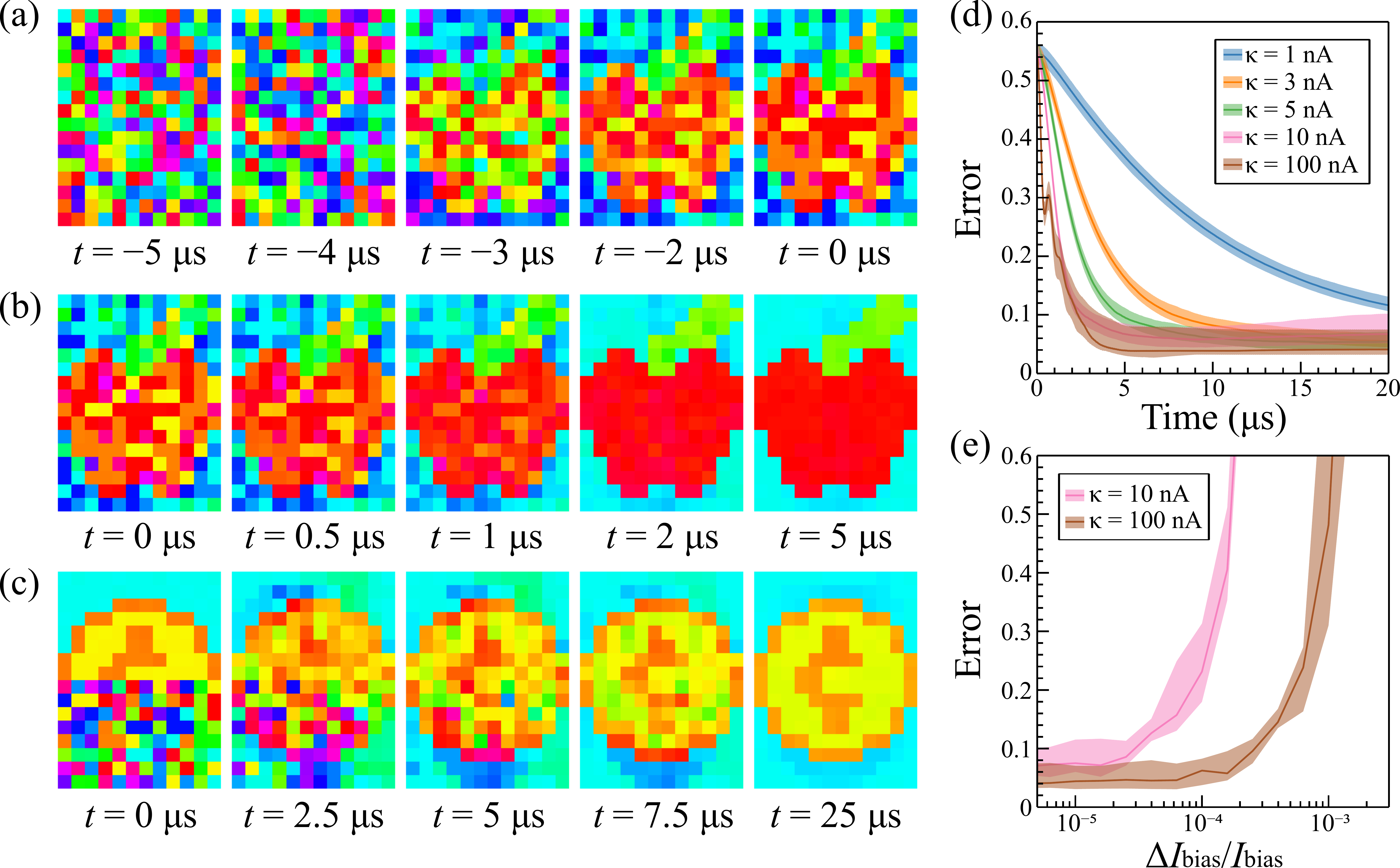}
    \caption{Representative snapshots of the image represented on the oscillator array (a) in the preparation stage and (b) in the retrieval stage. The entire query image is distorted using random, discrete noise that gives a root mean square distortion of one state along the color wheel. (c) Retrieval process of an incomplete query image. The upper-half the query image is intact, and the lower-half has randomized pixel values. (d) Error evolution as a function of time for 30 Gaussian distorted images of each of the 12 images, as illustrated in one distortion of one image in (b). (e) Average error at the end of 20~{\textmu}s, for 30 distorted images of each of the 12 images, plotted as a function of relative standard deviation of the Gaussian DC bias spread of the oscillators. In (d) and (e), the solid line corresponds to the average error and the translucent fills around each solid line are the quartile bounds. The different curves are for different strengths of the feedback current, $\kappa$, as described in Eq.~\eqref{eq:Ifeedback}. In (b) and (c), $\kappa=10$~nA.}
    \label{fig:time_evolution}
\end{figure}

\section{Network simulations}
\label{sec:simulations}

We simulate the performance of the network described in Sec.~\ref{sec:STNOcHopfield} assuming ideal performance of the feedback network for the synapses and the behavior of the spin-torque oscillators described in \ref{sec:appA}. The total current injected into each oscillator $j$ is a sum of the DC bias current $I_j^\mathrm{DC}$ and the AC synaptic feedback current $I_j^\mathrm{AC}$. $I_j^\mathrm{DC}$ sets the nominal oscillation frequency of the gyrotropic mode of the oscillator, chosen here to be 248~MHz by setting $I_j^\mathrm{DC} = 80$~{\textmu}A. On the other hand, $I_j^\mathrm{AC}$ is a sum of synaptic currents $I_{ij}^\mathrm{Syn}$ from the feedback network that feed into the $j$th oscillator. The total AC feedback into the neuron $j$ is given by
\begin{equation}
    I_j^\mathrm{\mathrm{AC}} = \sum_{i = 1}^{N} I_{ij}^\mathrm{Syn} = \sum_{i = 1}^{N} \kappa|w_{ij}|\sin\left[2\pi ft + \arg\left(\hat{a}_i\right) + \arg\left(w_{ij}\right)\right].
    \label{eq:Ifeedback}
\end{equation}
Here $\kappa$ is a coupling constant that controls the strength of the injected feedback current.

Figs.~\ref{fig:time_evolution}(a) and (b) show representative snapshots at various labeled time intervals during the two stages of the retrieval cycle, preparation and recognition. In Fig.~\ref{fig:time_evolution}(a), the oscillators are prepared to represent the query image. The color of each pixel represents the phase of the oscillator representing that pixel. The query image stored on the oscillator array is distorted using a discrete Gaussian noise generator, which distorts each pixel by one state on average from its nominal value. In the recognition phase of Fig.~\ref{fig:time_evolution}(b), the prepared image state evolves in the presence of the feedback network. The oscillator network relaxes to the stored  image of Fig.~\ref{fig:dataset}(a) that the noisy version of the image most resembles. 

Similarly, when presented with an incomplete image, the network can retrieve the complete image. The recognition stage of such a retrieval cycle is shown in Fig.~\ref{fig:time_evolution}(c). The oscillator array is prepared to represent an image with half the pixels representing an original image from the dataset in Fig.~\ref{fig:dataset}(a), while the other half of the pixels are independently assigned one of the twelve allowed values from a uniform  distribution of the color wheel. While the coupling to the disordered pixels pulls some of the correct pixels away from the correct value, taken as a whole, the set of pixels continuously evolves toward the correct image. 

We characterize the error between a retrieved image and the $k$th stored image as
\begin{equation}
    \Delta^k = \min_\phi \sqrt{\frac{1}{N}\sum_{j=1}^{N}\left(\arg\left\{\exp[i(\phi^k_{j}-\phi_{j}+\phi)]\right\}\right)^2} ,
\end{equation}
where $\phi_{j}$ is the phase of the $j$th pixel in the retrieved image and $\phi^k_{j}$ is the corresponding phase in the stored image.
This complicated expression is simply describing the root mean square difference in the phases between the stored and retrieved images.  The complications arise from the facts that the relative phases must be between $-\pi$ and $\pi$ and that the overall relative phase does not matter as shown in Fig.~\ref{fig:dataset}(c). The argument of the exponential function of a complex argument keeps the phase difference in the appropriate range and the minimization over $\phi$ shifts the relative phases to get the best agreement.

Figure~\ref{fig:time_evolution}(d) shows the evolution in the error between a set of distorted images and the correct images averaged over 30 distorted versions of each of the twelve images shown in Fig.~\ref{fig:dataset}(a). At time $t=0$~{\textmu}s, the oscillators are prepared to be in one of the distorted images  using a Gaussian noise generator on each image of the dataset that distorts each pixel on average by one discrete phase level, $2\pi/12$, from its nominal value. The average error decreases as the oscillator dynamics brings the oscillators into one of the fixed points representing the correct stored image. Each color represents a different coupling constant $\kappa$. Stronger coupling constants lead to a quicker initial decrease in the error, but this improved rate of convergence largely saturates for values of $\kappa$ greater than 10~nA. We note that the value of $\kappa$ is the upper bound on the feedback associated with each pair of neurons. When summed over all other neurons, typical values of the total feedback current for $\kappa=10$~nA are around 150~nA, ranging up to close to 1~{\textmu}A for some neurons.

The primary assumption used in the analysis thus far is that all oscillators run at a set fixed frequency. While it is possible to achieve phase and frequency locking of oscillators in the presence of small frequency variations between the oscillators, large variations cause desynchronization resulting in a failed retrieval process. Fig.~\ref{fig:time_evolution}(e) shows the average error after 5~{\textmu}s of relaxation for 30 distorted versions of each of the 12 images in the dataset, as described for Fig.~\ref{fig:time_evolution}(d). These simulations are performed for two coupling constants $\kappa$. The oscillator frequencies are varied by varying the DC biases. The final error is plotted against the fractional variations of the DC biases, with each oscillator bias varied independently using a Gaussian distribution. When the oscillator array desynchronizes, the final state error is high. Higher coupling values result in a larger threshold for desynchronization because the locking range increases linearly with the amplitude of the AC feedback current for values near or below 1~{\textmu}A. A bias tolerance of 0.01~\% or lower is required for successful retrieval for a coupling of $\kappa = 10$~nA. For oscillators with a distribution of parameters, this tolerance indicates the degree to which the DC bias currents need to be controlled to keep the oscillators synchronized. While the sensitivity to variation in the locking frequencies could be improved by increasing the AC feedback amplitude, the improvement comes at the cost of increasing the energy consumed in the calculation.

\begin{figure}
    \centering
    \includegraphics[width=0.9\textwidth]{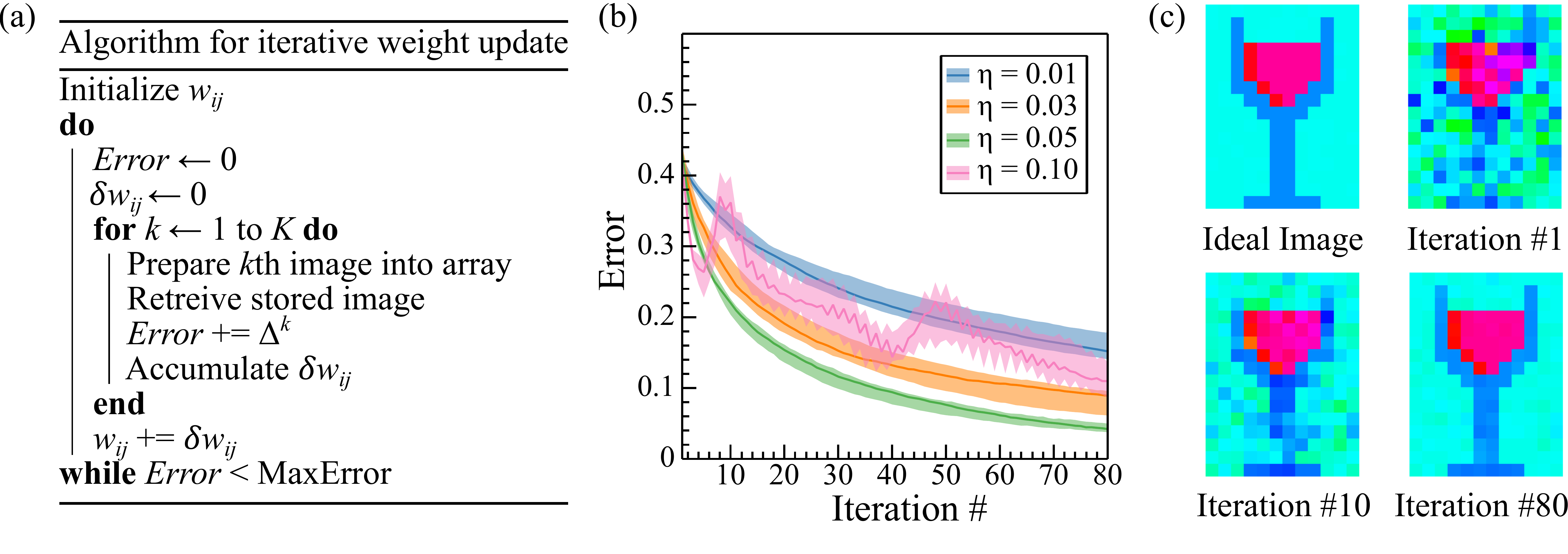}
    \caption{(a) Iterative weight update procedure (b) Error as a function of iteration number for four different values of the learning rate parameter $\eta$. The initial weight matrix elements are assumed to be implemented in the synaptic matrix with an error of 20~\% each in both magnitude and phase. The solid line corresponds to the average error and the translucent fill around each solid line are the quartile bounds. (c) Ideal image and examples of a retrieved image of a wine glass at the end of first and tenth and eightieth weight update iterations for $\eta=0.05$. While snap shots of only one image are shown, these are captured while the weights for all images were updated simultaneously.}
    \label{fig:iterative}
\end{figure}

The simulations thus far assume perfect implementation of offline-trained weights. In the presence of nonideal writing, layout-related delay differences between synapses, and device-to-device variations, the synapses may produce a different value of delay and scaling than intended. These errors can be iteratively corrected to ensure proper storage of images on the oscillator array. 

A typical iterative procedure to correct the implemented weights is shown in Fig.~\ref{fig:iterative}(a). With the initial weights implemented into the synaptic feedback matrix, the oscillator array is prepared to represent one of the $K$ original images with pixel values $\hat{x}_i^k$. Starting from this stored image state, the network is left to evolve in the recognition stage configuration. In the presence of imperfections, the network relaxes into an incorrect image with pixel values $\hat{a}_i^k$. This incorrect image retrieval process starting from ideal images is repeated for the entire dataset while accumulating the error and the weight update corrections. Here, we use a contrastive divergence learning rule, which gives the weight update $\delta w_{ij}$ correction at the end of each iteration as
\begin{equation}
    \delta w_{ij} = \frac{\eta}{N} \sum_{k=1}^{K} [\hat{x}_i^k (\hat{x}_j^k)^* - \hat{a}_i^k (\hat{a}_j^k)^*], \label{eq:contrastive_divergence}
\end{equation}
where $\eta$ is the learning rate. The weight update is carried out at the end of each iteration provided the total accumulated error is greater than the acceptable tolerance. 

An example of the reduction of error with every weight update iteration is shown in Fig.~\ref{fig:iterative}(b) for four different learning rates $\eta$. The weight matrix elements are prepared to be in a state with a normal error of 20~\% each in their magnitude and phase. Weight updates are deterministic and are performed using the weight update rule in \eqref{eq:contrastive_divergence} at each iteration. Notice that the error reduces with the increase in the number of iterations, with the rate of reduction of error controlled by $\eta$. The rate of reduction of error increases concomitantly with $\eta$ until a critical value is reached. Beyond this critical value the corrections become too large to iteratively correct towards the right solution, as is the case with $\eta=0.10$. Representative snapshots of retrieved images for $\eta = 0.05$ at the end of first, tenth, and eightieth iterations are shown in Fig.~\ref{fig:iterative}(c) along with an ideal image for comparison.

\section{Circuits for complex synaptic weights}
\label{sec:circuits}

\begin{figure}
    \centering
    \includegraphics[width=1.0\textwidth]{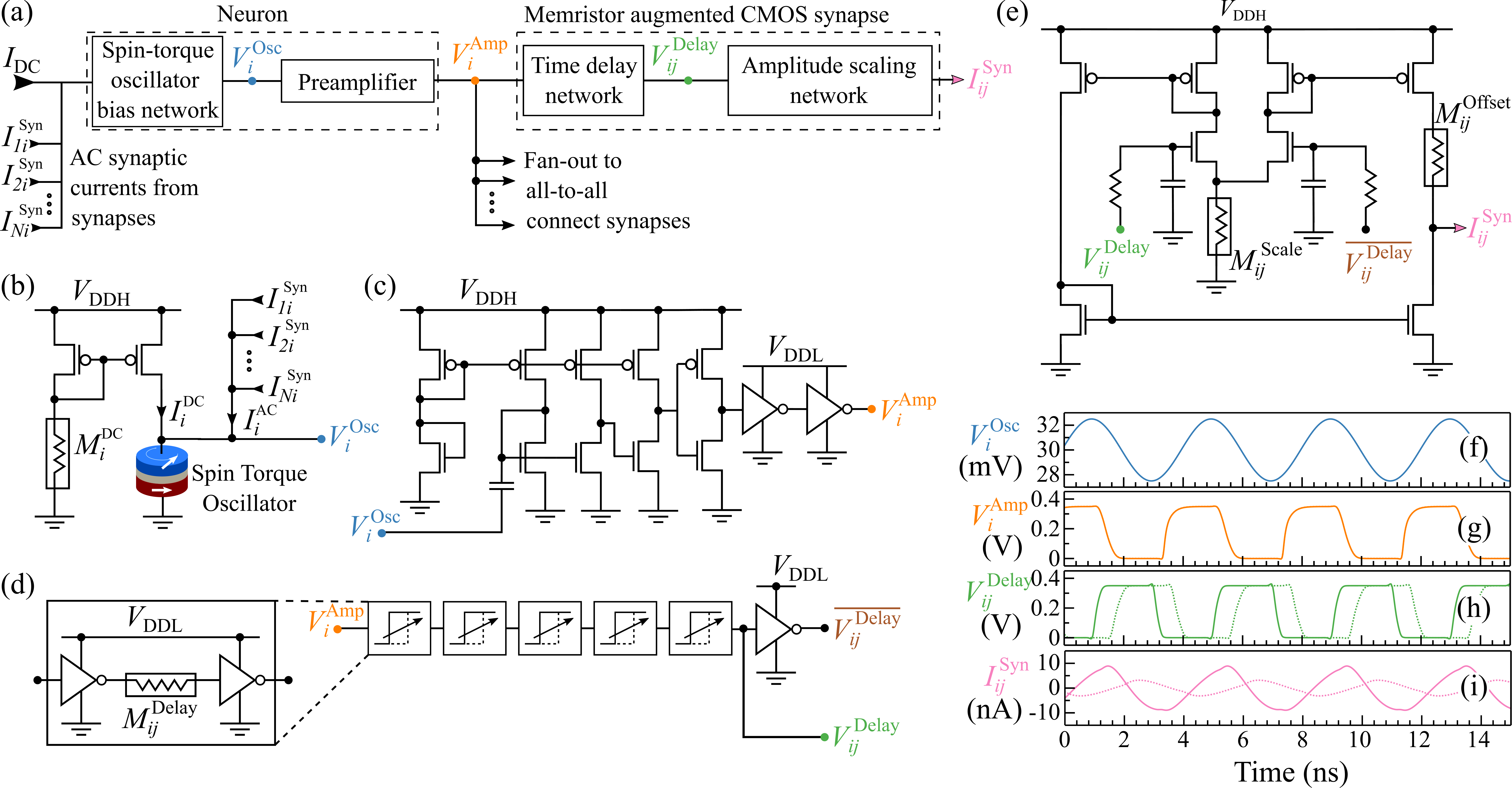}
    \caption{CMOS circuit implementation of a spin-torque oscillator- and memristor-based Hopfield network. (a) Unrolled feedback circuit showing the connections of the blocks detailed in the rest of the figure. The voltages and currents at the output of each block are plotted in panels (f)--(i). (b) The spin-torque oscillator bias network that controls the DC current through the oscillator to cause it to oscillate at the chosen frequency. The low voltage synaptic input currents $I_{ij}^\textrm{Syn}$ modify the phase of the oscillations. The resistance variations associated with the oscillations convert part of the DC input current $I_i^\textrm{DC}$ into an oscillatory output voltage $V_i^\textrm{Osc}$ that is much larger than the voltage oscillations due to the synaptic input currents. (c) Preamplifier. The low voltage AC input is converted to a square wave output ($V_i^\textrm{Amp}$) \change{with a lower line voltage of $V_\textrm{DDL}$ when compared to the nominal line voltage of $V_\textrm{DDH}$}. (d) Time delay network. \change{A series of delay elements convert the input square wave to a phase-shifted output voltage $V_{ij}^\textrm{Delay}$ and its complement $\overline{V_{ij}^\textrm{Delay}}$}. (e) Amplitude scaling network. The memristor $M_{ij}^\textrm{Scale}$ controls the conversion of the input square wave to the small amplitude AC output current $I_{ij}^\textrm{Syn}$, which is then fed back into the input of the appropriate neuron. (f)--(i) Voltage wave forms. The colors of the curves correspond to the values labeled in colored symbols in the other panels. (f) Assumed voltage variation across one of the oscillators. (g) Output of the preamplifier, which amplifies the oscillator output by roughly a factor of 100. (h) Voltage output of the delay circuit for two different delay values (solid and dotted). (i) Obtained synaptic currents to be injected into two other neurons with different scale factors in addition to the previously applied delays. }
    \label{fig:circuit}
\end{figure}

This section describes the detailed circuitry required to realize a complex Hopfield network using spin torque oscillators. The network is composed of neuron and synapse circuits which are recurrently connected as shown in Fig.~\ref{fig:concept_circuits}(c). Unrolled versions of the blocks involved are shown in Fig.~\ref{fig:circuit}(a), each of which is discussed in more detail below. The neuron circuit consists of a bias network,  which sets the current flowing through the spin-torque oscillator, a summation circuit, which sums the synaptic currents from the various branches, and a preamplifier circuit, which generates a square wave from the output voltage of the spin torque oscillator. This square wave passes through the synapse circuitry, which takes this square wave output and passes it through a programmable time delay network that performs phase shifting, and then a programmable amplitude scaling network that generates a scaled synaptic current output. These currents are fed back into the appropriate neurons, as the recurrent loop is completed. The output of each of the $N$ neurons feeds the input of a synapse connecting to each of the other $N-1$ neurons, then each neuron sums the input from the other $N-1$ neurons. 

Detailed versions of the neuron circuits are shown in Fig.~\ref{fig:circuit}(b) and (c). The bias network consists of a memristor, $M_i^\textrm{{DC}}$ in Fig.~\ref{fig:circuit}(b), in series with a diode-connected p-channel metal-oxide semiconductor field effect transistor (MOSFET). That pair is connected in parallel with another p-channel MOSFET in series with the spin-torque oscillator. Since both of those p-channel MOSFETs see the same gate-source voltage and are the dominant resistances in the two lines, the memristor-controlled current from the first branch is mirrored into the spin-torque oscillator. If the spin-torque oscillators could be fabricated with sufficient uniformity, see Fig.~\ref{fig:time_evolution}(e), a single programmable current branch could be shared amongst them to minimize power consumption. In the likely event that such uniformity is too difficult to fabricate, the memristors for each oscillator need to be tuned to give the same oscillation frequency for all of the oscillators to within 0.1~\% depending on the feedback current [see Fig.~\ref{fig:time_evolution}(e)]. The AC synaptic currents from the various synapses are summed into each oscillator by wiring them together. The oscillators respond to this summed current by adjusting their phase. Since the amplitude of the oscillation does not vary significantly, the DC current through the spin-torque oscillator produces an output voltage, $V_i^\textrm{Osc}\approx5$~mV as shown in  Fig.~\ref{fig:circuit}(f). This output signal is much smaller than $V_\textrm{DDH}=$ 0.7~V.

The output signal is converted into a square wave by a preamplifier circuit shown in Fig.\ref{fig:circuit}(c). \change{No information is lost through this conversion of the oscillator signal into square wave because only the oscillation phases contain  information and the phase information is preserved when the signal is converted into a square wave}. Operating on square waves significantly reduces the power consumption in the delay circuitry in the synapses because the circuit only draws current during the transitions. A capacitor AC-couples the output of the oscillator to an appropriate bias voltage. 
The first part of the preamplifier has five branches connected to the nominal supply voltage in this technology, $V_\textrm{DDH}$. In the first branch, both transistors are diode connected and increased in size relative to the minimum at the chosen technology node to setup a bias current, which is mirrored into the second branch. The second branch uses this bias current to generate the appropriate bias voltage for the next two branches which form a two-stage amplifier. The third and fourth branches are connected in a common source topology, in which the n-channel MOSFET behaves as the input gain element, while the p-channel MOSFET behaves like a current source. The last branch, which is an inverter, produces a full-swing (0~V to $V_\textrm{DDH}$) square wave output. In order to achieve sufficiently high gains, the transistors forming the gain stages of the preamplifier are sized to be twice the minimum length for the particular technology model we use to simulate the circuits. The final part of the preamplifier consists of two appropriately sized inverters operating at a reduced supply voltage $V_{\mathrm{DDL}}=0.35$~V to match the synaptic circuit. The output is fanned out to the synaptic elements. The output of the preamplifier, denoted by $V_i^\textrm{Amp}$, is shown in Fig.~\ref{fig:circuit}(g).

A key feature of this network is the energy efficiency of the synaptic circuits that allow the implementation of complex weights for the feedback currents. The synaptic circuits consist of a time delay network and an amplitude scaling network as shown in Figs.~\ref{fig:circuit}(d) and (e). The time delay network is composed of a series of five delay cells. These delay cells, shown in the inset of Fig.~\ref{fig:circuit}(d), consist of two low voltage inverters and a memristor. The charging time of the gate of the second inverter of the delay cell is programmable through the $RC$ time constant set by the gate capacitance $C$ and the effective resistance $R$ which consists of the MOSFET channel resistances and the tunable resistance of the memristor $M_{ij}^\textrm{{Delay}}$, thus determining the delay of each cell. \change {The delay elements operate at a reduced supply voltage $V_\mathrm{DDL}$. The active power in the delay networks scales as $V_{DDL}^2$. Therefore, a lower $V_{DDL}$ restricts the charging currents minimizing the power consumption of the delay network.} Combining multiple delay elements ensures that the delays can be tuned to allow phase shifts between 0 to $2\pi$ as determined by $M_{ij}^\textrm{Delay}$. 
An additional inverter is included to output the complementary signal as needed in the amplitude scaling network.

\change{Additional parasitics in the time delay network, such the capacitance of the memristor, would increase the energy cost of producing delays while also affecting the range of achievable delays. However, typical memristor parasitic capacitances are a fraction of the gate capacitances used in this study \cite{fouda2017modeling, madhavan2018high} and we estimate that the achievable delays per RC block of the time delay network decrease by less than fifteen percent. Such reductions in the achievable delay ranges could be readily compensated by the addition of additional delay blocks or by allowing a larger resistance values for each memristor.}


The scaled synaptic current is generated by feeding the inverted and non-inverted outputs of the delay cells to a differential  transconductance amplifier, as shown in Fig.~\ref{fig:circuit}(e). The differential transconductance amplifier provides a scaled current output proportional to the applied differential voltage, with the scaling set by the memristor $M_{ij}^\textrm{Scale}$. The output of the delay cells is low-pass filtered by an $RC$ network which sets the bias point of the input common mode to $V_\mathrm{DDL}/2$. The memristor $M_{ij}^\textrm{Scale}$ controls the total currents in both branches and hence sets the gain of the input stage. The p-channel and the n-channel MOSFETs that comprise the current mirror steer the copied currents until they are subtracted from each other at the output branch. An offset tuning memristor $M_{ij}^\textrm{Offset}$ is added to account for drain-source voltage mismatches between the right and left branches. Since only a differential current is being output by the circuit, its magnitude can be kept low, while the individual branches of the circuit still operate at a larger current, constrained by the bias currents required for correct circuit functionality. \change{The bandwidth of the scaling circuit designed here is much higher than the fundamental frequency of the oscillator. When we included the parasitic capacitances of the memristors, the changes to the feedback current were small enough that we did not observe any noticeable change in the circuit operation.}

The oscillator phases are most affected by frequency components of $I^\mathrm{AC}_{i}$ that are close to the oscillator's set frequency. Feedback components not in the range of the oscillator's locking frequencies, in particular, higher harmonics, have a weaker effect on the oscillator dynamics as long as their amplitude is low. We have tested the oscillator model we use by simulating the injection of sine waves, triangle waves, and square waves. All give similar locking ranges. We conclude from these simulations that for the size of the input currents we consider, $I^\textrm{Syn}_{ij}$ in Fig.~\ref{fig:circuit}(i), the higher harmonics do not affect the performance of the oscillator.  

\begin{figure}
    \centering
    \includegraphics[width=1.0\textwidth]{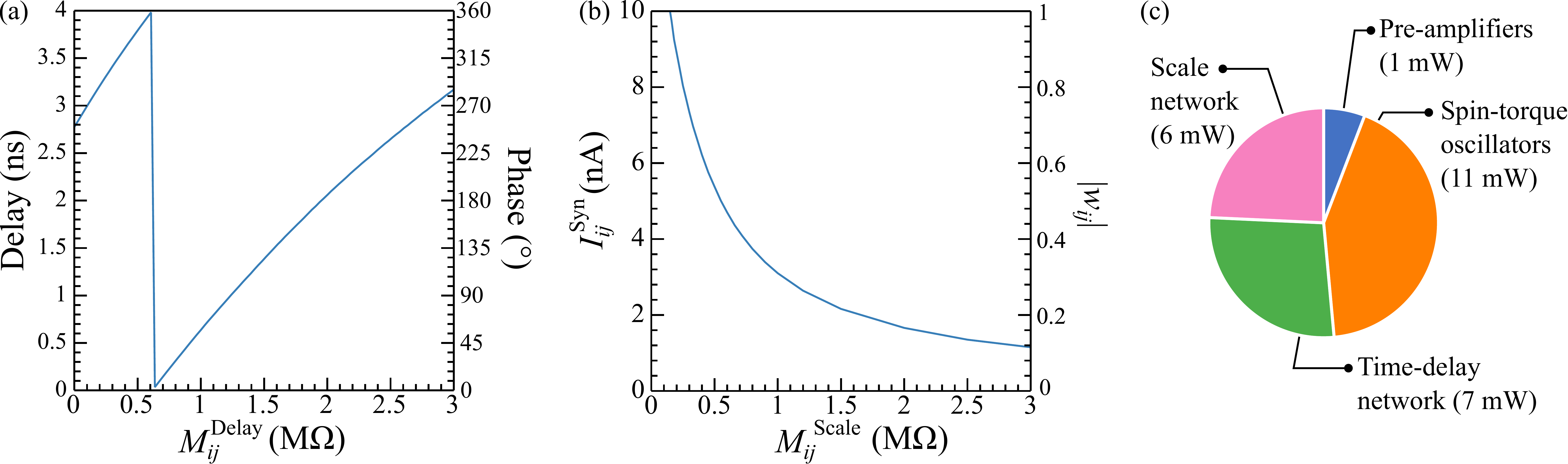}
    \caption{(a) Delay obtained from the delay network compared to the oscillator output (left axis) and the equivalent phase (right axis) versus the delay memristor resistance $M_{ij}^\mathrm{Delay}$. (b) The injected synaptic current amplitude (left axis) and its equivalent scaling (right axis) versus the scaling memristor resistance $M_{ij}^\mathrm{Scale}$. (c) Power consumption of various elements of the implemented complex Hopfield network with 192 spin-torque oscillators.}
    \label{fig:delay_scaling_power}
\end{figure}

These CMOS circuits were designed using the Predictive Technology Model 16~nm high-power technology~\cite{arizonaPTM} and were simulated in Cadence Virtuoso.\footnote{Certain commercial products or company names are identified here to describe our study adequately. Such identification is not intended to imply recommendation or endorsement by the National Institute of Standards and Technology, nor is it intended to imply that the products or names identified are necessarily the best available for the purpose.} They either operate at the nominal supply voltage for this technology, $V_\textrm{DDH} = 700$~mV, or a reduced supply voltage, $V_\textrm{DDL} = 350$~mV. Using a reduced supply voltage helps achieve the desired range of delays while reducing the power consumption. The memristors in the circuit are treated as simple resistors. \change{Programming the memristors requires circuitry in addition to what we describe here, as we discuss in Sec.~\ref{sec:discussion}.}

Representative waveform outputs measured at various nodes labeled in Fig.~\ref{fig:circuit}(a) are shown in Figs.~\ref{fig:circuit}(f)--(i). These waveforms are measured for an unrolled network, without feedback. We assume a 248~MHz sine wave output for a spin-torque oscillator, as seen in Fig.~\ref{fig:circuit}(f), and track the intermediate nodal voltages and the output synaptic current of a synapse connected to this spin-torque oscillator. Fig.~\ref{fig:circuit}(g) shows the saturated square output of the preamplifier. The preamplifier drives a time-delay network and a dummy inverter load that is sized to emulate the large fan-out at the end of the preamplifier. The time-shifted output voltages of the time-delay circuit for two chosen delays are shown in Fig.~\ref{fig:circuit}(h). The waveforms are square waves with a full voltage swing phase shifted by the selected phases. Finally, Fig.~\ref{fig:circuit}(i) shows the synaptic current output of the scale network that scales the two time-delayed waveforms in Fig.~\ref{fig:circuit}(h) by two different chosen values of scaling. Notice that each circuit in Figs.~\ref{fig:circuit}(b)--(e) introduces intrinsic constant phases to the oscillator output. The physical layout of this network will cause additional delays associated with the different path lengths. However, these intrinsic phase shifts can be compensated by appropriately adjusting the delays produced by the synaptic delay elements. 


The synaptic weights are stored in memristors that control the phase shifting and the scaling of the feedback current. Fig.~\ref{fig:delay_scaling_power}(a) shows, as a function of the memristor resistance $M_{ij}^\mathrm{Delay}$, the measured delay (left axis) produced at the end of the delay network relative to the spin-torque oscillator output. The right axis gives the phase corresponding to this delay. As a result of the intrinsic delays introduced in other parts of the circuit, the zero-delay case occurs for a non-zero value of $M_{ij}^\mathrm{Delay}$. Notice that a complete range of phase shifts from 0$\degree$ to 360$\degree$ can be produced for a resistance range of approximately 20~k$\Omega$ to 3 M$\Omega$.
Fig.~\ref{fig:delay_scaling_power}(b) shows the measured synaptic current (left axis) and its corresponding weight scale factor (right axis) plotted against the memristor resistance $M_{ij}^\mathrm{Scale}$. The range of realizable scale factors is limited by the possible range of resistances. Furthermore, the variation of the synaptic current with respect to $M_{ij}^\mathrm{Scale}$ is highly non-linear, and as a result, for the training process, a look-up table is essential to correctly program the value of $M_{ij}^\mathrm{Scale}$ for a desired scaling value.

For the technology we simulate, Fig.~\ref{fig:delay_scaling_power}(c) gives the power consumed by various components of the 192-oscillator complex Hopfield network in Fig.~\ref{fig:circuit}(a). The 192 spin-torque oscillators along with the bias network draw about 11~mW of power and the corresponding preamplifier circuits following these oscillators draw about 1~mW of power. Although the power dissipation in individual synapses is small compared to the spin-torque oscillator networks, the large number of synapses ($192\times 191$) results in a significant power dissipation in the synaptic network. In the networks that we have designed, we estimate 7~mW of power consumption across all the time-delay networks and an additional 6~mW of power across the scaling networks. These estimates suggest an estimated 25~mW power consumption for this 192-spin-torque-oscillator- and memristor-based implementation of the complex Hopfield network. Assuming a 5~{\textmu}s inference time following the results in Fig.\ref{fig:time_evolution}(d), we estimate 130~nJ of energy consumed in the network per image retrieval. These power estimates do not include the energy spent in the preparation phase, which add roughly 55~nJ, nor does it include the energy needed to extract the oscillator phases for subsequent processing. The oscillations of the spin-torque oscillators could be extracted with the preamplifiers, making a negligible contribution to the energy. The cost of processing these phase fronts would depend on how the subsequent processing would be done.




\section{Discussion}
\label{sec:discussion}

We have theoretically demonstrated inference on a spin-torque-oscillator-based complex Hopfield network with 192 oscillators. The goal of this paper is to establish the requirements on device performance and reproducibility that would enable the implementation of this and related networks. \change{While the CMOS circuitry we use can be readily fabricated, the spin-torque oscillators and memristors we simulate are beyond the current state of the art.} In this section, we discuss some of the current limitations that need to be addressed. Hopefully, this discussion can guide future development efforts designed at taking advantage of spin-torque oscillators and memristors.

There are several obstacles to the immediate implementation of this approach. The first is the ability to synchronize a large number of spin-torque oscillators. Current experimental demonstrations of electrically synchronizing spin-torque oscillators are limited to a few oscillators with synchronization times limited to a few milliseconds~\cite{tsunegi2018scaling}. \change{Such experimental demonstrations are performed in small-scale research labs that lack large scale production capabilities. Commercial fabrication facilities could have the needed capabilities to produce sufficiently uniform devices, however, no attempts have been made to do so. Therefore, } implementing large arrays of spin-torque oscillators for phase-based computing remains \change{unexplored}. Strict control over the frequency dispersion of the individual oscillators, which could be achieved by refining the quality of materials during fabrication, is essential to increase the number of oscillators that can be synchronized. For the low feedback currents considered here, synchronizing 192 oscillators would require a frequency dispersion of less than 0.1~\% as shown in Fig.~\ref{fig:time_evolution}(e). \change{The upper bound on the feedback coupling strength is set by non-linearities associated with the oscillators. As the summed feedback currents become comparable the bias current, the phase locking characteristics of the oscillator to a mixture of ac signals is no longer linear. Additional distortions may be introduced by the CMOS feedback network with increase in the feedback strength. Smaller oscillator networks allow for larger feedback strengths which correspond to a higher tolerance of frequency dispersion.}

\change{One aspect that is important to investigate before realizing oscillator networks in practice is noise.} Spin-torque oscillators operating at finite temperatures have phase noise~\cite{keller2009time,dussaux2012field}, which will interfere with synchronization. These measurements report resonance peak widths that are 0.1~\% to 1~\% of their free running frequencies. These widths are broader than the frequency spreads required by the network described above. However, synchronization with an external source, here, the feedback from the other oscillators, significantly reduces the phase noise. Whether this finite temperature phase noise requires greater feedback than is used in this network is left to future work. 

\change{The CMOS components used in the feedback circuitry also introduce noise. This contribution can be particularly significant to the feedback currents that are in the range of a few nanoamperes. We performed SPICE simulations including thermal noise arising from the CMOS scaling network and determined that the thermal noise amplitudes have an root mean square value of 15 nA, thus being comparable to the feedback currents arising from the scaling network. However, because the oscillator locking range is restricted to a megahertz or less depending on the strength of the feedback, the presence of a broadband noise such as that arising from thermal sources has little effect on its frequency locking behavior, as we have checked in simulations of the oscillators. Consequently, we expect that CMOS noise has little impact on the network performance provided feedback currents are not made too small.}

Another obstacle is programming the memristors used to provide local memory in the network. \change{We require analog control of memristor resistances to encode the complex weights and to control the bias currents of the individual oscillators in the case that oscillators need to be individually tuned to match their frequencies. Local analog memory using memristors is an active area of research \cite{wang_resistive_2020, sarwat_phase-change_2022} and our approach would not be viable without its realization.  There are different constraints on the memristors we use in different parts of the network. While errors in precise analog control of memristors used for programming the complex weights only introduce errors in recalled vectors, errors in programming the bias control memristors result in catastrophic decoherence of the oscillator networks.}

We have not included memristor programming and control circuitry in Fig.~\ref{fig:circuit}. While memristors with the chosen resistance ranges have been fabricated~\cite{xia2019memristive,li2018analogue}, such memristors require high-voltage thick oxide transistors to decouple the low-voltage circuits from  larger [up to $\approx$2~V] forming and write voltages. Designing such circuits would require either working in a technology that includes these higher voltages and associated transistors or the development of memristors with improved capabilities. Improving the voltage-level compatibility of memristors with highly scaled CMOS transistors is an important goal of ongoing research~\cite{golonzka2019non, jain201913, yao_fully_2020} for almost all envisioned applications of memristors. \change{Without sufficient reduction in memristor writing voltages, the overhead associated with the write circuitry and the associated energy costs could be far higher than that of the inference circuitry used in this study.}

If it were possible to fabricate oscillators with a sufficiently low frequency spread, this paper shows that low power CMOS circuitry can be used to couple them in ways to enable energy-efficient processing. Identifying paths to increase the efficiency of this approach is complicated by the combination of technologies involved. The power shown in Fig.~\ref{fig:delay_scaling_power}(c) is balanced between the neuronal and synaptic parts of the circuit but for larger numbers of neurons will be dominated by the synaptic parts of the network that scale as the number of oscillators squared. As the size of images is increased, the power will scale as the square of the size of the stored images. However, the energy per inference may scale more steeply than that if it takes longer for the larger network to converge. It may be possible to reduce the power consumed by the synapses by using a more advanced technology node, but scaling arguments are not simple. Unlike scaling arguments for pure CMOS circuits, the properties of the spin-torque oscillators and the memristors must scale in the right way as well.

Power consumed by the spin-torque oscillators could be reduced further by scaling down the dimensions of the oscillators. Again, the scaling arguments are not simple for several reasons. Larger current densities are required to sustain oscillations as spin-torque oscillators are scaled down~\cite{dussaux2012field}. As a result, although the total DC current is reduced by scaling down the oscillators, this reduction is lower than the area reduction obtained by scaling. On the other hand, scaling down the oscillators increases the oscillation frequency. Use of faster spin-torque oscillators reduces the synchronization time, and thus would be expected to reduce the image retrieval times and the associated overall energy costs. Faster oscillations also allow the design of synapse circuitry with lower $RC$ time constants, further driving down the power dissipated. Current experimental realizations of spin-torque oscillators have been limited to diameters of hundreds of nanometers~\cite{khalsa2015critical, tsunegi2016self}, as presumed in this study. Theoretically predicted scaling suggests that the spin-torque oscillator radius could be reduced to tens of nanometers~\cite{romera2018vowel}. As discussed in~\ref{sec:appA}, the rate of phase convergence depends on the strength of the field-like torque, so using materials that increase this parameter will lead to faster convergence and a lower energy cost.

Non-linear oscillators could be implemented using alternative technologies such as spin-Hall oscillators~\cite{chen2016spintorque}, vanadium-dioxide oscillators~\cite{corti2020scaled} and opto-electronic oscillators~\cite{farhat85optical}. The delay and scaling networks proposed in this work can be tuned to operate between tens of megahertz to a gigahertz. Slower oscillations require large $RC$ time constants, thus increasing the power consumption. Therefore, alternative methods to realize tunable delay elements would be needed for slow oscillators such as those based on vanadium-dioxide. Frequencies higher than a gigahertz would be possible but would require redesigning the circuits at the cost of increasing the power required. However, we consider it likely that the energy per computation would be similar or lower because the power increase would be offset by a decrease in computation time resulting from a higher operating frequency.

There has been significant recent progress in developing more powerful Hopfield networks by implementing energy functions with terms with higher powers of the neuron states than quadratic~\cite{ramsauer2021hopfield, krotov2021large}. The quadratic model used here allows for all-to-all connections for all pairs of neurons. Implementing higher order terms in the energy functional electrically would require many neuron interactions and exploding levels of circuitry. For example, a quartic term would generate for each oscillator contributions to the feedback from all possible trios of oscillators, increasing the number of synaptic pathways from the roughly $4\times 10^4$ considered here to $16\times 10^8$. The difficulty in wiring such many-body interactions reflects the complications involved in treating many-body interactions in physical systems. It may be possible to make approximate implementations modeled after some of the successful approximations from many body physics. 

The all-to-all connections in this type of network make it most suitable for storing information in which each ``pixel'' is strongly correlated with each other one. One possible way to improve the performance of Hopfield networks for image processing could be to take advantage of the locality of information in images. In existing implementations, each pixel is connected equally strongly to all other pixels no matter how close they are spatially. Neural networks~\cite{lawrence1997face} frequently use local convolutional filters in the initial layers to process pixels that are spatially close. A similar approach, such as making the interactions in the energy functional local, might significantly reduce the energy consumption without substantially reducing the performance of the network. For other applications, reducing the range of interactions might lead to more efficient implementations without loss of performance if the range is tuned to match the range of correlations in the stored data. 

This section discusses several different lines of development needed for the hardware implementations of Hopfield networks proposed here. A key enabler will be advances in manufacturing uniformity and reliability for both spin-torque oscillators and memristors as these technologies transition from laboratories to fabrication facilities. Our results show that it is possible for hardware implementations of Hopfield networks to encode multilevel or continuous information, the possibility of which greatly increases the types of data that could be usefully stored in such an array.






\section{Acknowledgements}
The authors thank Jabez McClelland, Matthew Daniels, Matthew Pufall, and William Borders for critical readings of the paper.
Nitin Prasad designed the network and performed the simulations of the circuit and is supported by Quantum Materials for Energy Efficient Neuromorphic Computing, an Energy Frontier Research Center funded by the U.S. DOE, Office of Science, Basic Energy Sciences, under Award DE-SC0019273. Advait Madhavan, who acknowledges support under the Cooperative Research Agreement Award No. 70NANB14H209, through the University of Maryland, and Prashansa Mukim designed and simulated the CMOS circuitry. All authors discussed the results and wrote the paper.

\appendix 
\section{Spin-Torque Oscillator Model}\label{sec:appA}

There have been several types of spin-torque oscillators that have been experimentally characterized~\cite{kaka2005mutual, pribiag2007magnetic}. For specificity, we focus on a vortex-based spin-torque oscillator based on a circular MTJ in which the magnetization of the free layer assumes a vortex configuration~\cite{pribiag2007magnetic}. Under the influence of the spin-polarized tunneling current, the resulting spin transfer torque causes the vortex center to precess around the center of the disc.  As it does, the relative orientation between the magnetization in the precessing vortex layer and the fixed layer changes giving rise to an oscillating resistance.

We model the zero-temperature dynamics of the free-layer of the vortex-based spin-torque oscillator using a Thiele approach~\cite{guslienko2008magnetic}. In this approach, the dynamics of the vortex state is parameterized by the polar coordinates ${\bf r}  = r_0 (\rho \cos\theta, \rho\sin\theta,0)$ of the vortex center in the plane of the free layer, where $r_0$ is the radius of the disc, making the radial coordinate $\rho$ dimensionless. This approach captures the dominant {\em gyrotropic mode} of the vortex and ignores vortex distortions. The time evolution of the vortex is described by~\cite{khalsa2015critical}
\begin{align}
    \dot{\rho} & = a\rho - b\rho^{3} - c \cos(\theta),  \nonumber \\
    \dot{\theta} & = \omega_0 + \omega_1\rho^2 +\frac{c}{\rho}\sin(\theta),\label{eq:ModelBasic}
\end{align}
where
\begin{align}
    a & = \frac{a_\mathrm{J} J_i}{G} - \frac{D_0}{G}\omega_0, \nonumber\\
    b & = \frac{D_1}{G}\omega_0+\frac{D_0}{G}\omega_1, \nonumber\\
    c & = \frac{b_\mathrm{J}J_i}{G}, \nonumber \\
    \omega_0 & = \frac{1}{G}\left(\kappa_\mathrm{MS}^0 + \kappa_\mathrm{Oe}^0 J_i\right), \nonumber\\
    \omega_1 & = \frac{1}{G}\left(\kappa_\mathrm{MS}^1 + \kappa_\mathrm{Oe}^1 J_i\right). \label{eq:ModelParameters}
\end{align}
The description and values corresponding to the geometry- and material-dependent parameters that appear in Eq.~\eqref{eq:ModelParameters} are given in Table \ref{tab:ModelParameters}. The nominal bias current through the $i$th spin-torque oscillator $I_i^\mathrm{DC} = J_i^\mathrm{DC}(\pi r_0^2)$ is set to 80~{\textmu}A so that the spin-torque oscillator produces self-sustained oscillations at 248~MHz. The total injected current $I_i = I_i^\mathrm{AC} + I_i^\mathrm{DC}$ (and the corresponding total current densities $J_i = J_i^\mathrm{AC} + J_i^\mathrm{DC}$), where $I_i^\mathrm{AC}$ is the sum total AC feedback from all synaptic connections. 

\begin{table}[]
\centering
\caption{Spin-torque oscillator parameters used in the study.}
\begin{tabular}{llrl}
\hline
\textbf{Parameter}     & \textbf{Description}                            & \textbf{Value}         &                        \\ \hline
$G$                    & Gyrovector amplitude                            & $1.14 \times 10^{-13}$ & $\mathrm{Js/(m^2 rad)}$ \\
$D_0$                  & $1^\mathrm{st}$-order damping constant                    & $5.08 \times 10^{-16}$ & $\mathrm{Js/(m^2 rad)}$ \\
$D_1$                  & $2^\mathrm{nd}$-order damping constant                   & $9.51 \times 10^{-17}$ & $\mathrm{Js/(m^2 rad)}$ \\
$\kappa_\mathrm{MS}^0$ & $1^\mathrm{st}$-order magnetostatic confinement constant  & $1.41 \times 10^{-4}$  & $\mathrm{J/m^2}$       \\
$\kappa_\mathrm{MS}^1$ & $2^\mathrm{nd}$-order magnetostatic confinement constant & $3.53 \times 10^{-5}$  & $\mathrm{J/m^2}$       \\
$\kappa_\mathrm{Oe}^0$ & $1^\mathrm{st}$-order Oersted field confinement constant  & $3.40 \times 10^{-16}$ & J/A                    \\
$\kappa_\mathrm{Oe}^1$ & $2^\mathrm{nd}$-order Oersted field confinement constant & $-1.70 \times 10^{-16}$ & J/A                    \\
$a_\mathrm{J}$         & Orthogonal spin-transfer efficiency parameter   & $3.10 \times 10^{-16}$ & J/A                    \\
$b_\mathrm{J}$         & In-plane spin-transfer efficiency parameter     & $8.26 \times 10^{-17}$ & J/A                    \\
$r_0$                  & Free-layer radius                               & $100$                  & nm       \\  \hline           
\end{tabular} \label{tab:ModelParameters}
\end{table}

In the network simulations in Sec. \ref{sec:simulations}, we integrate these equations to obtain the time evolution of the vortex cores of the oscillators. The oscillators evolve to a synchronized state determined by their AC feedback currents. To connect Eq.~\eqref{eq:ModelBasic} to the formulation used in the main text in terms of the complex number $\hat{a}_i=e^{i\phi_i}$ encoding the phase, we write $\theta_i=\omega t + \phi_i$. The time dependence of $\hat{a}_i$ is then
\begin{equation}
    \frac{d\hat{a}_i}{dt} = i\hat{a}_i ( \dot{\theta}_i - \omega ) . \nonumber
\end{equation}
The DC current is adjusted for each oscillator so that in the absence of an AC current, $\omega=\omega_0+\omega_1 a/b$ is the chosen operating frequency, where for each oscillator $\omega_0$, $\omega_1$, $a$, and $b$ take values in the DC limit. The time evolution of $\hat{a}_i$ depends on the instantaneous values of $\theta$ and $\rho$ and the input AC component of the current through a complicated function determined by Eq.~\eqref{eq:ModelBasic}. Unfortunately, we have not been able to identify a limit in which this function becomes physically transparent.

We can gain insight into the phase-locking process by focusing on low order effects. First, consider the case with no AC current so that all parameters in Eq.~\eqref{eq:ModelBasic} are time independent. Neglecting higher harmonic behaviors, we assume
\begin{align}
    \rho&= \rho_0 +\rho_1 \cos(\omega t + \chi) \nonumber\\
    \theta&= \omega t + \phi + \phi_1 \cos(\omega t + \zeta ) . \nonumber
\end{align}
The steady state contributions in Eq.~\eqref{eq:ModelBasic} lead to
\begin{align}
    \rho_0^2 &= a_0 / b \nonumber\\
    \omega &= \omega_0 + \omega_1 \rho_0^2 = \omega_0 + \omega_1 a_0 / b , 
    \label{eq:steadystate}
\end{align}
where $a_0$ is the part of $a$ due to just the DC current. There are additional contributions to this equation from the static part of quantities like $\rho_1^2 \cos^2(\omega t + \zeta )$, but they can be neglected because $\phi_1$ and $\rho_1$ are small.
The equation of motion for $\rho$ becomes
\begin{align}
    - \omega \rho_1 \sin(\omega t + \chi) &= a_0 \rho_1 \cos(\omega t + \chi) - 3 b \rho_0^2 \rho_1 \cos(\omega t + \chi ) - c \cos( \omega t + \phi ) , \label{eq:distortion}
\end{align}
where we neglect the $\phi_1 \cos(\omega t + \zeta )$ inside the $\cos$ because it leads to higher harmonics and static contributions that are higher order in $\phi_1$. Using the static results and noting that $a_0 \ll \omega$ gives $\rho_1 \approx c / \omega $, which is small because $c\ll\omega$, and $\chi\approx\phi+\pi/2$. After cancelling the static parts and neglecting higher harmonics, the equation of motion for $\theta$ becomes
\begin{align}
     - \omega \phi_1 \sin(\omega t + \zeta ) &= 2 \omega_1 \rho_0 \rho_1 \cos(\omega t + \chi) + \frac{c}{\rho_0} \sin( \omega t + \phi ) . \nonumber
\end{align}
Substituting $\pi/2+\phi$ for $\chi$ and $c/\omega$ for $\rho_1$, collecting terms, and simplifying gives $\phi_1\approx{c}/{(\omega\rho_0)}$ and $\zeta=\phi+\pi$. This fixes all parameters of steady precession in the absence of AC currents except for the arbitrary phase $\phi$.  Adding in the AC current fixes $\phi$ relative to the phase of the AC input. Numerical simulations consistently give $\phi=\pi/2$ in the phase-locked state. The small values we find for $\delta_1$ and $\theta_1$ mean that the precession is very circular in this regime and the AC voltage output very sinusoidal.

These results indicate which parameters in Eq.~\eqref{eq:ModelBasic} determine different aspects of the motion. The steady state results, Eq.~\eqref{eq:steadystate} show that the damping-like torque, through $a$, and the damping, primarily through $b$, determine the steady state radius of gyration, $\rho_0$ and the frequency $\omega$. The frequency-dependent contribution, which distorts the circular orbit, as seen in $\rho_1$ and $\phi_1$ in Eq.~\eqref{eq:distortion}, enters through the field-like torque parameter, $c$. With no AC current present, the overall phase of the oscillation, $\phi$ is undetermined. Introducing an AC current adds an AC driving term through $a$ into Eq.~\eqref{eq:ModelBasic}. The only terms in Eq.~\eqref{eq:ModelBasic} that depend on the phase of the gyration are those derived from the field-like torque. Thus, the strength of the field-like torque determines how quickly the phase locks to the phase of the AC driving current.

Assuming that the fixed layer is magnetized in-plane along the $x$-direction, the variation resistance of the spin-torque oscillator as a function of the oscillating core coordinates is
\begin{equation}
    \Delta R = \lambda \xi \Delta R_0 \rho \sin(\theta). 
\end{equation}
Here, $\lambda$ is the average magnetization to vortex displacement ratio, taken to be 2/3, $\xi=\pm1$ depending on the helicity and the direction of oscillation of the vortex core. $\Delta R_0 = (R_\mathrm{AP}-R_\mathrm{P})/2$, with $R_\mathrm{AP}$ and $R_\mathrm{P}$ being the resistance of the MTJ when the free layer is magnetized entirely antiparallel and parallel, respectively, with respect to the fixed layer.

\section*{References}

\providecommand{\newblock}{}


\end{document}